\documentclass[useAMS,usenatbib,referee]{mn2e}

 \usepackage{Times}

\usepackage{graphicx}

\usepackage{natbib}
\bibpunct{(}{)}{;}{a}{}{,}

\usepackage[english]{babel}

\title[Non-detection of excited nascent H$_{2}$ in dark
clouds]{Laboratory evidence for the non-detection of excited nascent
H$_{2}$ in dark clouds}
\author[E. Congiu et al.]
 {E. Congiu,$^{1}$\thanks{E-mail: emanuele.congiu@u-cergy.fr} E.~Matar,$^1$
  L.~E.~Kristensen,$^1$\thanks{present address: Leiden Observatory, Niels Bohrweg 2, NL-2333 CA Leiden, The Netherlands}
  F.~Dulieu,$^1$
  and J. L. Lemaire$^1$ \\
$^{1}$Universit\'e de Cergy-Pontoise \& Observatoire de Paris,
LERMA/LAMAp, UMR 8112 du CNRS, 95000 Cergy-Pontoise, France}

\begin{document}

\date{Accepted 2009 May 22. Received 2009 May 21; in original form 2009 February 23}

\pagerange{\pageref{firstpage}--\pageref{lastpage}} \pubyear{2009}

\maketitle

\label{firstpage}

\begin{abstract}
There has always been a great deal of interest in the formation of
H${_2}$ as well as in the binding energy released upon its formation
on the surface of dust grains. The present work aims at collecting
experimental evidence for how the bond energy budget of H${_2}$ is
distributed between the reaction site and the internal energy of the
molecule. So far, the non-detection of excited nascent H${_2}$ in
dense quiescent clouds could be a sign that either predictions of
emission line intensities are not correct or the de-excitation of
the newly formed molecules proceeds rapidly on the grain surface
itself. In this letter we present experimental evidence that
interstellar molecular hydrogen is formed and then rapidly
de-excited on the surface of porous water ice mantles. In addition,
although we detect ro-vibrationally excited nascent molecules
desorbing from a bare non-porous (compact) water ice film, we
demonstrate that the amount of excited nascent hydrogen molecules is
significantly reduced no matter the morphology of the water ice
substrate at 10 K (both on non-porous and on porous water ice) in a
regime of high molecular coverage as is the case in dark molecular
clouds.

\end{abstract}



\section{Introduction}

The formation of molecular hydrogen (H${_2}$) in the interstellar
medium (ISM) is considered one of the most important chemical
reactions occurring in space. H${_2}$ is ubiquitous, it is by far
the most abundant molecular species in the Universe and is a major
contributor to the cooling of astrophysical media. In addition, this
simple molecule is also responsible for initiating the interstellar
chemistry leading to the large and various inventory of molecules
that have been observed so far \citep{dal00}.

Since gas phase routes for the conversion of hydrogen atoms to
molecules are too inefficient to account for the high abundances
that are observed, it has been long assumed \citep{gou63}, and now
confirmed by many laboratory experiments \citep[e.g.,][and
references therein]{vid06}, that molecular hydrogen is formed
efficiently in surface reactions on cosmic dust grains. The two main
mechanisms invoked for surface catalysis are: the
Langmuir-Hinshelwood (L-H) mechanism, in which H${_2}$ forms in a
H-atom diffusion process, and the Eley-Rideal (E-R) mechanism, or
`prompt' mechanism, in which an impinging H-atom reacts directly
with an adsorbed H-atom \citep{dul96}.

Of major concern for this work, is the 4.48~eV released upon
formation of the H${_2}$ molecule on grain surfaces and how the
binding energy is distributed between the reaction site and the
translational and internal energy in the molecule. Several
theoretical and experimental works have been carried out on this
subject. To cite just a few examples, the formation of H${_2}$ via
the E-R process on graphite surfaces at 10~K was studied with purely
quantum mechanical calculations \citep{far00,mei01,mor04b},
semi-classical quantum molecular dynamics \citep{rut01} and
classical and quasi-classical trajectory calculations \citep{par98}.
These studies generally find the product H${_2}$ in significantly
excited \textsl{v--J} states. Quantum dynamics of the L-H mechanism
on graphite was studied by \citet{mor04a,mor05}. They predict even
greater vibrational energy of the nascent molecule. A very recent
computational study by \citet{gou09} on an olivine surface seems to
suggest, however, that nascent H$_{2}$ formed from chemisorbed
H-atoms is considerably less ro-vibrationally excited than when it
is formed on graphite.
 As to the case of interest here,
hydrogen formation on water ice, \citet{tak99} made a detailed
calculation of H${_2}$ formation on amorphous ice using a classical
description in a molecular dynamics formulation. They found most of
the recombination energy distributed in vibration (70~to 80\%) of
the product H${_2}$ while only 5\% of the H-H bond energy was
deposited in the ice substrate.

Experiments performed on non-porous surfaces show that molecular
hydrogen is formed in a ro-vibrational excited state both on
graphite \citep{lat08} and on amorphous solid water ice
\citep{ami07}. As for the formation of H${_2}$ on porous substrates,
\citet{ros03} and \citet{hor03} found that newly formed HD molecules
desorb from a porous water ice film with a kinetic energy that shows
previous thermalization with the substrate. However, their
experiments do not provide any information on the internal
excitation of the molecules upon formation.

In principle it is possible to observe newly formed molecular
hydrogen directly in the interstellar medium. The formation
excitation of ro-vibrational states will be followed by spontaneous
radiative transitions to lower energy levels hence infrared emission
from newly formed H$_2$ in dark cores should be detectable
\citep{dul93b}. Observations of Barnard 68 (or Lynds Dark Nebula LDN
57) have been made as part of the programme ref 66.C-0526(A) in
March 2001 at ESO-VLT using the ISAAC instrument in spectroscopic
mode by Lemaire \&~Field (private communication). No detection of
H$_2$ emission has been observed looking at the core of the object
in K band, despite a 5-hour integration in nodding mode to subtract
the telluric lines. The position of the slit on the object was
chosen on the core and off object including, in the latter case, a
few faint stars in or on the side of the spectrograph slit. Then
only the continuum for these non-listed stars has been detected.
 This lack of detection has
been reported later on by \citet{tin03} on two others dark nebulae,
L1498 and L1512, using the CGS4 spectrometer at UKIRT and observing
in the H band. Using the calculations by \citet{far00} and
\citet{mei01}, \citet{tin03} predicted an observational spectrum
arising from newly formed excited H$_2$. According to their model,
there should be a clear spectral signature of H$_2$ formation
excitation in both diffuse and (particularly) dark clouds. They then
went on to observe two dark clouds to verify their predictions.
H$_2$ emission was not detected at the 3$\sigma$ level. The authors
provide several reasons for this non-detection. Possibly one of the
most important reasons is that the predicted spectrum is based on
dust grain surfaces composed of pure graphite. In the case of the
L1498 and L1512 it is known that molecular species such as CO, CS,
H$_2$CO and CCS have been depleted from the gas phase and reside on
the surface of the grains \citep[e.g.,][and references
therein]{shi05}. It is therefore very likely that grains are covered
in molecular ices. The dust temperature is estimated to be
$\sim10$~K in both clouds. The other main assumption is that the
formation of H$_2$ proceeds through the E-R mechanism although it is
likelier that the formation proceeds through the L-H mechanism
\citep[e.g.,][]{cre06}.

This letter is intended to provide another strong argument to
explain the non-detection of excited product H${_2}$ in dark clouds.


\section[]{Experimental Methods}\label{sectexp}

The experimental technique and apparatus have been described
elsewhere \citep{ami06} and are only summarized here. The
experiments take place under ultra-high vacuum (UHV) conditions
(base pressure $\sim 1 \times 10^{-10}$~mbar) in a stainless steel
chamber. This experiment directs a D-atom beam collimated through a
triply differentially pumped beam-line to a target located in the
vacuum chamber. The target consists of a polished circular (1~cm in
diameter) copper surface attached to the cold finger of a
closed-cycle He cryostat and can be cooled to a lower limit of 8~K.
The temperature is measured by a calibrated silicon diode clamped on
the back of the target. D-atoms generated by microwave dissociation
of D$_2$ are piped through a PTFE tubing to an aluminium nozzle in
the first stage of the beam-line. A typical dissociation efficiency
in this experiment was 60\% and the resulting D-atoms were
thermalized to approximately 300~K via collisions with the walls of
the PTFE tube.
Amorphous solid water (ASW) ice samples were grown \textit{in situ}
by exposing the cold Cu surface to pure water vapor by back-filling
the chamber. Non-porous (compact) ASW (np-ASW) ices were prepared at
a surface temperature of 120~K while porous ASW (p-ASW) samples were
prepared with the surface held at 10~K. The preparation and
characterization methods are the same as employed in a recent work
by our group \citep{fil09}.

A quadrupole mass spectrometer (QMS) is employed for the detection
of the products entering the vacuum chamber or coming off the sample
in real-time mode during the irradiation phase. Due to the
configuration of our apparatus, the QMS was placed 5~cm above the
surface to allow for direct D-irradiation of the ASW ice sample. The
QMS is also used to detect D$_{2}$ molecules in a vibrationally
excited state by tuning the ionizing electron energy inside the QMS
head as follows. The QMS detection is preceded by ionization of the
molecules that are then selected by their \textit{m/e} ratio and
finally detected as ion counts. The ionization threshold for the
lower level of D$_2$ is 15.46~eV and the ionizing electron energy is
normally set to 30~eV. As the energy requirement for ionization
decreases for the vibrationally excited molecule, setting the
electron energy to a value lower than 15.46~eV allows to detect
solely internally excited molecules. In the present experiment, the
detection of excited D$_{2}$ was performed with the ionizing
electron energy set to 15.2~eV, which would allow us to detect
molecules in a vibrationally excited state $v'' \ge 2$. Previous
experiments carried out using this technique proved effective for
the detection of excited D$_2$ up to $v'' = 7$ \citep{ami07}. Newly
formed molecules are not detected directly as they leave the ASW
sample but as background molecules that have collided at least once
with the walls of the vacuum chamber. However, it can be fairly
assumed that this configuration has the only disadvantage of
reducing the number of detectable excited molecules but that no
qualitative effect occurs.

Nascent excited D${_2}$ molecules were studied by D-atom irradiation
of a 100-monolayer (1~ML $= 1.0 \times 10^{15}$ molecules cm$^{-2}$)
np-ASW ice and D-atom irradiation of a 10~ML p-ASW ice film grown on
a substrate of np-ASW. In both cases the substrate temperature was
held at 10~K.


\section[]{Results}\label{sectres}

\subsection{D irradiation of non-porous ASW}

\begin{figure}
 \resizebox{\hsize}{!}{\includegraphics{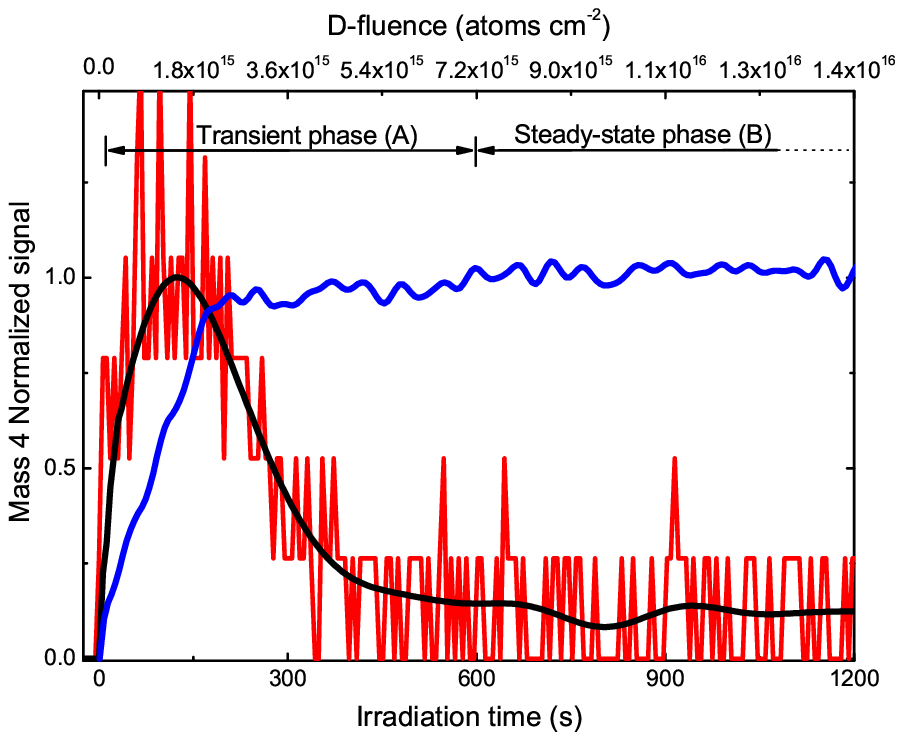}}
\caption{Normalized total-D${_2}$ signal (blue curve) and
excited-D$_{2}$ signal (red curve) monitored during D-irradiation of
a np-ASW ice film at 10~K. The black curve is a 20-datapoint (120 s)
FFT smoothing of the excited-D$_{2}$ signal. See Section~\ref{md}
for full description of the transient and steady-state phases (A)
and (B).} \label{1}
\end{figure}

Fig.~\ref{1} shows a comparison between the two curves obtained by
monitoring the total amount of D${_2}$ molecules and only the
excited-D${_2}$ signal (two separate experiments) during a 1-hour
exposure of D-atoms to the np-ASW ice substrate. The blue trace
represents the signal from all D${_2}$ molecules, whether in an
excited state or not. As discussed in detail in \citet{ami07}, the
roughly linear increase of the total-D${_2}$ signal during the first
200~s of D-irradiation indicates an enhancement of D$_{2}$
formation. Because of the augmented number of molecules on the
surface, deposited by the non-dissociated fraction of the impinging
D beam, the D-atom sticking probability increases \citep{gov80}. In
the early stage of the D-atom irradiation, when the surface coverage
is still low, the hallmark of the ongoing molecular formation --
besides the increase of the total-D${_2}$ signal -- is the
simultaneous increase of the number of excited nascent D${_2}$
molecules detected (red trace in Fig.~\ref{1}). In such a coverage
regime we assume that the excited-D${_2}$ yield is proportional to
the total number of nascent molecules. This also agrees with the
results obtained on another non-porous substrate (highly orientated
pyrolitic graphite) by \citet{cre06} at a surface temperature of
15~K. As the number of adsorbed molecules on the np-ASW substrate
approaches the saturation coverage of $\sim 1.8 \times 10^{14}$
molecules~cm$^{-2}$ \citep{ami07}, the total-D${_2}$ trace becomes
less steep (i.e., decrease of the recombination rate) until it
reaches a plateau value after $\sim 600$~s. A steady-state regime
between formed/deposited and desorbed D${_2}$ molecules is now
established and the recombination rate remains constant.
Concurrently with the probable decrease of the recombination
probability we also observe a drop in the excited-D${_2}$ yield
after it peaks at $t \sim 130$~s and reaches rapidly ($t \sim
500$~s) a steady low-count signal that is about $1/5$ of the maximum
value. Together with a reduced recombination probability with the
onset of the saturation coverage, the decrease of the
excited-D${_2}$ signal suggests that newly formed D${_2}$ molecules
may also undergo a thermalization with the surface via an efficient
energy transfer due to the enhanced number of molecules on the
surface. The low-count plateau value of ro-vibrationally excited
D${_2}$ molecules is however not to be seen as a mere instrumental
noise, but as a real and measurable signal. In fact, much attention
has been paid to verifying that we do not observe excited D$_{2}$
molecules when the ASW ice substrate is exposed to a beam of
molecules (i.e., no D$_{2}$ formation in progress) nor when D-atoms
are exposed to a surface at a higher temperature (i.e., residence
time of atoms on the surface becomes too short for recombination to
be efficient).

\begin{figure}
 \resizebox{\hsize}{!}{\includegraphics{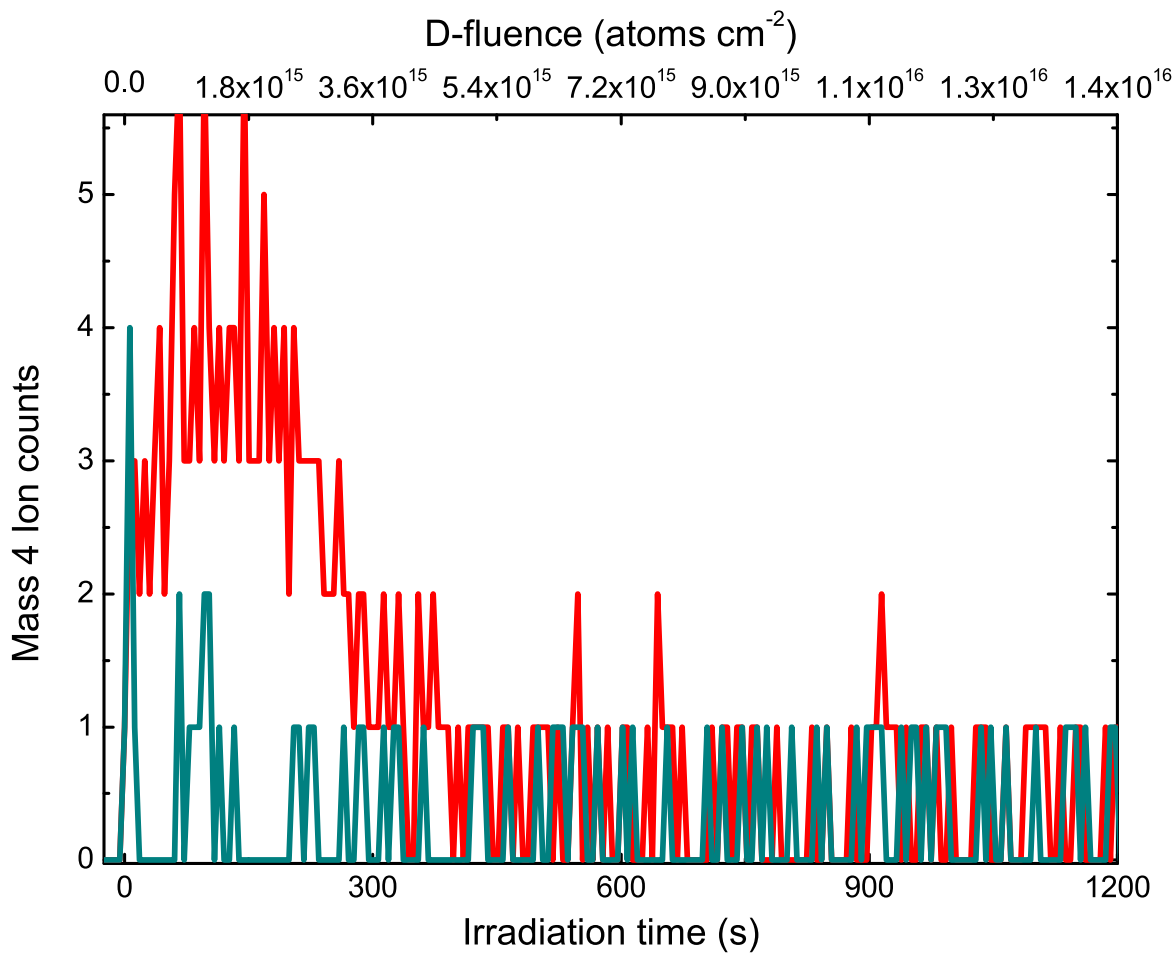}}
\caption{Excited-D${_2}$ signal monitored during D-irradiation of a
bare np-ASW ice film (red curve) and of a np-ASW ice surface
previously saturated with D${_2}$ molecules at 10~K (dark cyan
curve). The dwell time during recording of the excited-D$_{2}$ ions
was 6 seconds.} \label{2}
\end{figure}

In order to prove the decisive role of already adsorbed molecules in
a regime of saturation coverage and at what extent they determine
the drop-off in the amount of excited nascent D${_2}$ molecules, we
repeated the D exposure of the np-ASW substrate after dosing the
surface with D$_{2}$ molecules. To saturate the surface, we
irradiated the non-porous ice substrate to a 20-minute D${_2}$
exposure (corresponding dose of $\sim 10$~exposed ML). However, it
should be noted that when the saturation at 10 K is achieved, only a
fraction ($\sim 0.2$) of the non-porous water ice surface is covered
with molecules. This is due to a reduced sticking coefficient of
D${_2}$ molecules and a reduced binding energy. Fig.~\ref{2}
displays a comparison between the excited-D${_2}$ yields \textit{vs}
D-irradiation time of a bare np-ASW surface (red trace, case
discussed above) and of a np-ASW surface pre-dosed with molecules
(dark cyan trace). This figure clearly shows that the
excited-D${_2}$ yield in the case of a surface with D${_2}$
pre-adsorbed on the np-ASW film is much lower and a low-count
plateau is attained rapidly after the D-irradiation begins.

\subsection{D irradiation of porous ASW}

\begin{figure}
 \resizebox{\hsize}{!}{\includegraphics{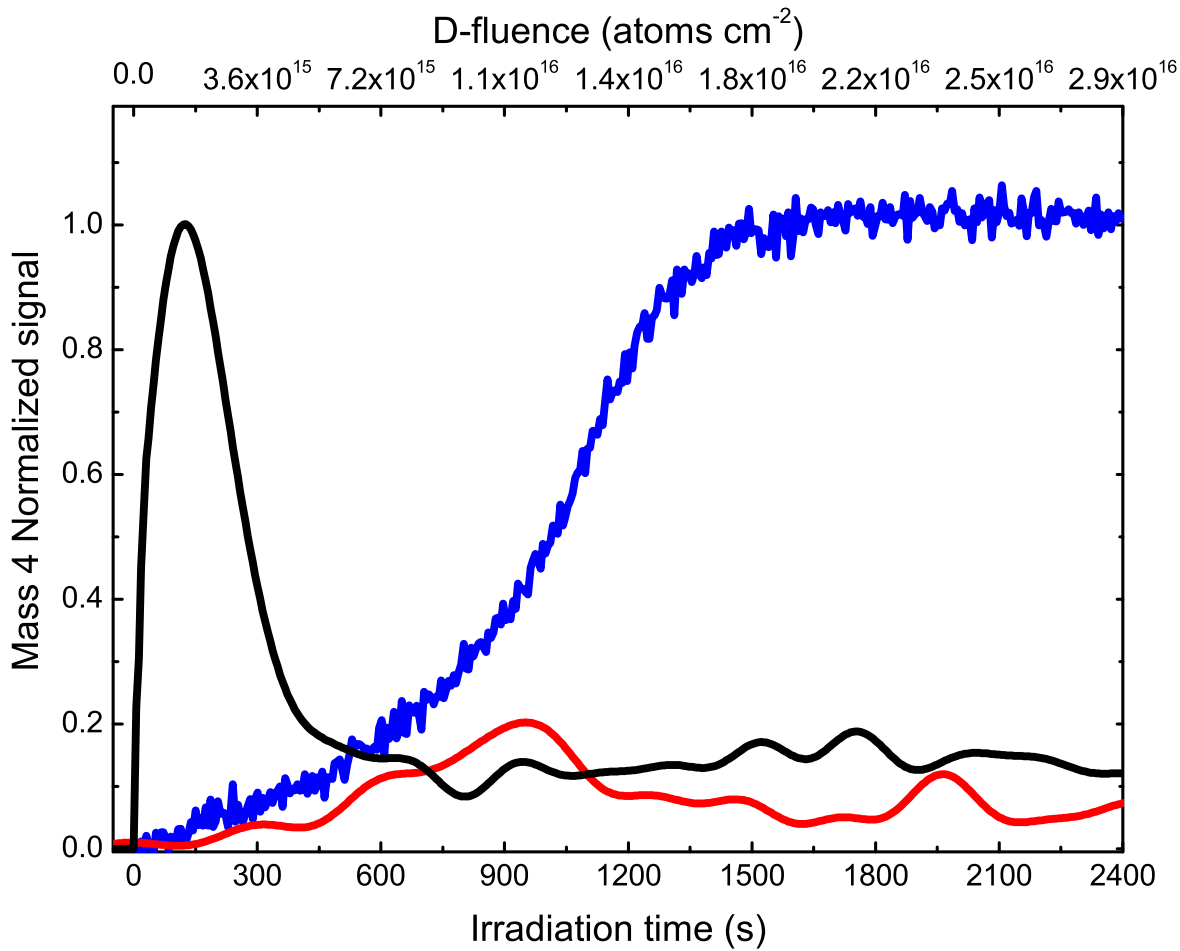}}
\caption{Normalized total-D${_2}$ signal (blue curve) and
excited-D${_2}$ signal (red curve) monitored during D-irradiation of
a 10-ML porous ASW ice film at 10~K. The black curve represents the
excited-D$_{2}$ signal monitored during D-irradiation of a np-ASW
ice film and is shown here for comparison of the excited-D${_2}$
yields in the np- an p-ASW case. Both black and red traces are
20-datapoint (120 s) FFT averages.} \label{3}
\end{figure}

In Fig.~\ref{3} we show the results obtained during D irradiation of
a 10~ML p-ASW ice film held at 10~K. The blue trace represents the
normalized total-D${_2}$ yield \textit{vs} D exposure time. Since
the effective surface area of the porous substrate is considerably
larger than that of the non-porous substrate,
we observe a slower completion of the saturation coverage (compare
with blue curve in Fig.~\ref{1}). As we have seen in the case of the
np-ASW substrate, the excited-D${_2}$ signal (red line) is likewise
proportional to the total yield of D${_2}$ molecules until a maximum
is reached after around 1000~s of D-atom irradiation, concurrent
with a fast rise of the molecular formation efficiency (slope of the
total-D$_{2}$ yield). This is in agreement with the assumption made
earlier that, in a regime of low coverage, the number of
excited-D${_2}$ molecules is proportional to the total number of
nascent molecules. However, the peak detected in the case of p-ASW
is so small that it barely stands out from the excited-D$_{2}$
low-count plateau value that is attained right after the maximum (at
$t \approx 1200$~s). To emphasize the big difference between the
amount of ro-vibrationally excited molecules detected in the two
types of water ice substrates investigated (np- \textit{vs} p-ASW),
the excited-D$_{2}$ yield curve of the porous substrate case (red
curve) and the excited-D$_{2}$ yield curve of the non-porous surface
case (black curve) are both shown in Fig.~\ref{3}. To facilitate a
direct comparison, the red curve (porous case) was scaled by the
same factor used for normalizing the black curve (non-porous case,
Fig.~\ref{1}). From this comparison, it is clear that the
excited-D$_{2}$ yield is considerably reduced in the case of the
p-ASW substrate. This result implies that the excited nascent
molecules readily thermalize with the substrate as they form in the
pores of the p-ASW film.

\section{Discussion}\label{sectdiscuss}

\subsection{De-excitation mechanisms on the surface} \label{md}

In order to make assumptions as to what kind of mechanism causes the
drop in the excited-D$_{2}$ yield at saturation coverage at 10~K, it
is worth discussing in more detail what happens when the sample of
np-ASW is exposed to D-atoms (Fig.~\ref{1}).
We can define two main phases during the irradiation of the ASW ice
surface; (A) a transient phase comprised between D irradiation at
$t=0$ and the beginning of the saturation coverage ($t \sim 600$~s),
and (B) a steady-state regime that is established at the saturation
coverage and that goes on as long as the D irradiation is running.
In the first part of the transient phase (A), after the irradiation
begins at $t=0$, few atoms are expected to stick although those that
are adsorbed on the D$_{2}$-free surface can occupy the deeper
binding sites. This results in a long residence time that, in turn,
favours recombination. As the irradiation goes on, the change in
surface coverage affects greatly the recombination efficiency
\citep{sch76,gov05}. Molecules gradually fill up the surface and the
binding energy of adsorbed atoms decreases gradually as the
molecules saturate the stronger binding sites \citep{hix92}. Yet,
this effect is compensated for by the increase of the sticking
probability of D-atoms. The excited-D$_{2}$ peak in Fig.~\ref{1}
indicates a high recombination probability resulting from an optimum
equilibrium between the increase of the sticking efficiency and the
shortening of the residence time of D-atoms on the surface. The
surface coverage at this stage is $\sim 1.0 \times 10^{14}$
molecules cm$^{-2}$. As pointed out by \citet{gov05}, around this
molecular coverage range we observe an abrupt change in the binding
energy of atoms, in that it begins to decrease as the surface
coverage grows further. As a result, the excited-D$_{2}$ signal
begins to drop rapidly as more molecules are deposited on the
surface and this may reflect, at least in part, a decrease of the
recombination efficiency. On the other hand, a large number of
D$_{2}$ are still likely to be formed in an excited state but
promptly release their internal energy on the surface via an
efficient energy transfer that involves the favourable mass ratio
(1-to-1) between newly formed molecules and the molecules adsorbed
in the vicinity \citep{sch76}. The already-adsorbed molecules can in
turn use this energy to desorb in a low- or non-excited state via a
non-thermal desorption process. At the onset of the steady-state
phase (B), the excited-D$_{2}$ yield has already attained a
low-count plateau and stays as such in a regime of saturation of the
surface, that is to say under conditions close to those of the dense
regions of the ISM (cf. \S~\ref{ap}) and of concern to us.

In the case of p-ASW, \citet{hor03} measured the kinetic energy of
HD molecules formed on the surface and found that they fully
thermalize with the porous surface before desorbing. In this work we
show that also the internal energy of newly formed D$_{2}$ molecules
that rapidly desorb is deposited almost completely in the porous
ice. In fact, hydrogen atoms are mobile at 10~K and penetrate the
porous structure \citep{mat08}, thus nascent molecules are
re-captured several times within the pores. This explains why we do
not observe significant internal or kinetic energy in the desorbing
molecules.

Unfortunately, we are not able to give an estimate of the fraction
of the bond energy deposited in the np-ASW ice substrate. Yet,
qualitative indications are that it is greater than what predicted
by \citet{tak99} (5\% of the entire formation energy) but comparable
to a value between 40\% and 60\% found by \citet{cre06} and
\citet{yab08}. This is consistent with the findings by \citet{sch76}
and with preliminary results obtained recently by our group in an
experiment studying the effects of non-thermal desorption induced by
hydrogen recombination that will be published in a forthcoming
paper.

\subsection{Astrophysical implications} \label{ap}

Our experiments demonstrate unequivocally that two conditions
contribute to the prompt de-excitation of nascent hydrogen molecules
formed on the surface of amorphous water ice. In the first
experiment, carried out on np-ASW, nascent D$_{2}$ molecules are
promptly de-excited as the surface coverage reaches the saturation
value at 10~K. In the experiment performed on p-ASW, we found that
the excited-D$_{2}$ yield was always low and basically independent
of the surface coverage, indicating that the roughness of the
surface contributes efficiently to the relaxation of newly formed
molecules at the surface. These results are relevant to the
formation of molecular hydrogen in the ISM and provide a plausible
explanation as to why excited nascent H$_{2}$ is very difficult to
observe in quiescent dark clouds.

We can actually assess if, under interstellar conditions of interest
here, surface coverage and surface morphology are comparable to
those simulated in our experiments. To estimate the surface coverage
on interstellar dust grain we have to verify that the lifetime
$\tau$ of a molecule on the surface is comparable to the time
interval $\Delta t$ between the arrival of two molecules on a single
dust grain surface; i.e., that the arrival rate balances the
evaporation rate. Let us consider the time interval between the
impacts of two molecules on the surface of a dust grain. Typical
conditions of the dense quiescent medium are $T_\mathrm{gas} \sim
T_\mathrm{grain} = 10$~K, density $n_{\mathrm{H}_{2}} = 10^{4}$
cm$^{-3}$, and for our calculation  we can approximate the dust
grains to spheres with average radius $r_\mathrm{grain} = 0.1\,
\umu$m. Then we have

\[
\Delta t = (n_{\mathrm{H}_{2}} \, \bar{v} \, A/4)^{-1}.
\]

$\bar{v}$ is the mean velocity of the molecules and $A$ is the grain
surface accessible to the molecules in the gas phase. Under dense
cloud conditions $\Delta t$ turns out to be of $\sim 6$ hours. The
residence time $\tau$ is expressed as $\tau = 1/\nu _{0} \,
\exp{(E_\mathrm{a}/T_\mathrm{s})}$. $\nu_{0}$ is the vibration
frequency of a molecule on the surface and represents the number of
attempts per second to evaporate ($\nu_{0}=10^{13}$~s$^{-1}$).
$E_\mathrm{a}$ is the adsorption energy in units of Kelvin and
$T_\mathrm{s}$ is the surface temperature. The surface temperature
in our case is 10~K so the residence time of H$_{2}$ molecules on
the grain surface depends critically on the adsorption energy.
\citet{ami07} derived the distribution of $E_\mathrm{a}$ from
D$_{2}$ TPD spectra in a saturation regime on np-ASW at 10~K. They
found that $E_\mathrm{a}$ ranges from about 400~to 700~K (33 to
59~meV) and estimated the saturation coverage to be \mbox{$\sim 1.8
\times 10^{14}$ molecules~cm$^{-2}$}. Putting $E_\mathrm{a}=400$~K
in the expression of $\tau$ (corresponding value for the lesser
bound molecules) we find a residence time of $\sim 6.5$ hours, and
thus $\Delta t \approx \tau$. By scaling $E_\mathrm{a}$ to a lower
value suitable for H$_{2}$, presumably comprised  between 360~and
650~K \citep{vid91}, we find a minimum value for $\tau$ of around
10~minutes which does not impair the validity of our conclusion as
the fraction of weakly bound H$_{2}$ ($E_\mathrm{a} < 400$~K) does
not account for more than a few percent of the total surface
population \citep[e.g.,][]{hix92}.

The surface molecular coverage on interstellar dust grains in dark
clouds can then be estimated to be in the range \mbox{$1.5 - 2
\times 10^{14}$ molecules cm$^{-2}$}, a value consistent with
previous estimates by \citet{gov80} and \citet{buc94} and comparable
to the surface coverage we attain in our experiments. In addition,
since dust grains residing in dark clouds are covered by icy mantles
of many monolayers dominated by water ice, as in the present work,
our results are relevant to the excited newly formed H${_2}$
molecules in dark clouds. If we assume a non-porous ice substrate
covering cosmic dust grains, we have shown that the fraction of
newly formed excited H$_{2}$ molecules is considerably reduced in
the gas phase. Should the surface of dust grains in dark clouds be
covered with a porous water ice mantle, the binding energy of the
H-H reaction is released on the surface even more efficiently owing
to the pores that trap the molecules long enough so that they
de-excite inside the pores before leaving the grain.

In dark clouds both `quenching mechanisms' are at play on the
surface of dust grains, namely, a high molecular coverage is present
on the icy mantles and a surface morphology that is probably an open
porous structure. Our results are even more remarkable if we
consider that the experiments were performed on a 10~ML p-ASW ice
substrate whereas icy deposits of 100 monolayers are typically
associated with dense cloud mantles. In the light of our results,
the observational predictions made by \citet{tin03} for the H$_2$
ro-vibrational emission line intensities should be revised and the
non-detection occurred in the searches by observing dark and
quiescent cores is now explained. Furthermore, the important role
played by the high molecular coverage in the decrease of the
observable internally excited H$_2$ leads us to suggest that these
results are relevant not only to water ice substrates, but also to
other types of substrates whenever H$_2$ becomes far more abundant
than H on the grain surface.

These experiments, performed under conditions close to those
encountered in dark clouds, demonstrate that the excess energy from
hydrogen formation distributed over vibrational states flows almost
completely into the grain surface and very little, if any, H${_2}$
ro-vibrational emission becomes detectable.


\section*{Acknowledgments}

We acknowledge the support of the national PCMI programme founded by
the CNRS, the Conseil Regional d'Ile de France through SESAME
programmes (contract I-07-597R), the Conseil G\'en\'eral du Val
d'Oise and the Agence Nationale de Recherche (contract ANR
07-BLAN-0129). EC in particular thanks the ANR for a post-doctoral
grant.


\bibliographystyle{mn2e}
\bibliography{bibliography}

\newpage

\vspace{0.5cm} \normalsize\noindent The definitive version will be
available at www.blackwell-synergy.com

\bsp

\newpage

\newpage

\newpage

%

\label{lastpage}

\end{document}